\documentclass[11pt,letterpaper]{JHEP3}
\usepackage{graphics}
\usepackage{epsfig}
\usepackage{amssymb}
\usepackage{stmaryrd}
\usepackage{amsmath}
\usepackage{amsfonts}
\usepackage{mathrsfs}
\usepackage{graphicx}
\usepackage{subfigure}

\def\be{\begin{equation}}
\def\ee{\end{equation}}
\def\ba{\begin{eqnarray}}
\def\ea{\end{eqnarray}}

\title{On the critical phenomena and thermodynamics of charged topological
dilaton AdS black holes}

\author{Ren Zhao\\
Institute of theoretical physics, Shanxi Datong University, 037009
Datong, China \email{zhao2969@sina.com}}

\author{Hui-Hua Zhao\\
Institute of theoretical physics, Shanxi Datong University, 037009
Datong, China\\
Department of Physics, Shanxi Datong University, 037009 Datong, China\\
\email{kietemap@126.com}}

\author{Meng-Sen Ma\\
Institute of theoretical physics, Shanxi Datong University, 037009
Datong, China\\
Department of Physics, Shanxi Datong University, 037009 Datong, China\\
 \email{mengsenma@gmail.com}}

 \author{Li-Chun Zhang\\
Institute of theoretical physics, Shanxi Datong University, 037009
Datong, China\\
Department of Physics, Shanxi Datong University, 037009 Datong, China\\
}

\abstract{In this paper, we study the phase structure and
equilibrium state space geometry of charged topological dilaton
black holes in $(n+1)$-dimensional anti-de Sitter spacetime. By
considering the pairs of parameters $(P\sim V)$ and $(Q\sim U)$ as
variables, we analyze the phase structure and critical phenomena of
black holes and discuss the relation between the two kinds of
critical phenomena. We find that the phase structures and critical
phenomena drastically depend on the cosmological constant $l$ (or
the static electric charge $Q$ of the black holes), dimensionality
$n$ and dilaton field $\Phi $.}

\begin{document}

\section{Introduction}

Black hole physics is a subject at the intersection of general
relativity, quantum mechanics and statistical physics and field
theory. This makes the subject receive a lot of attention. Black
holes have been used as the laboratory of many kinds of theories,
specially the thermodynamics of black holes plays an important
roles\cite{JDB,JDB1,JDB2,BCH,hawking,hawking1}. The thermodynamic properties of black holes have been
studied for several years, although the exact statistical
explanation of black hole thermodynamics is still lacked.

It shows that black holes also have the standard thermodynamic
quantities, such as temperature, entropy, even possess abundant
phase structures like hawking-Page phase transition\cite{hawking2}
and the critical phenomena similar to ones in the ordinary
thermodynamic system. What is more interesting is the research on
charged, non-rotating RN-AdS black hole, which shows that there
exists phase transition similar to the van der Waals-Maxwell
vapor-liquid phase transition\cite{chamblin,chamblin1,Lemos}.

Motivated by the AdS/CFT correspondence\cite{aharony}, where the
transitions have been related with the holographic
superconductivity\cite{gubser,hartnoll}, the subject that the phase
transitions of black holes in asymptotically anti de-Sitter (AdS)
spacetime, has received considerable
attention\cite{lemos2,sahay,sahay1,sahay2,kastor}. The underlying
microscopic statistical interaction of the black holes is also
expected to be understood via the study of the gauge theory living
on the boundary in the gauge/gravity duality.

The
studies\cite{rabin,rabin1,rabin2,rabin3,rabin4,rabin5,rabin6,majhi}
on the phase transition and critical phenomena of black holes in AdS
spacetime indicate the black holes are similar to the van der Waals
vapor-liquid system. The $(Q\sim U)$(where $Q$ is the static
electric charge, $U$ is the electrostatic potential on the horizon)
phase diagram of black holes in AdS spacetime is almost the same as
the $(P\sim V)$ phase diagram in van der Waals vapor-liquid system.

Among the gravity theories with higher derivative curvature terms,
the Gauss-Bonnet (GB) gravity has some special features and gives
rise to some interesting effects on the thermodynamics of black
holes in AdS space\cite{cai,cai1,kim,cai2,dutta,lala1}. The phase
structure of a GB-AdS black hole was briefly studied in
\cite{cai,dey}. And in the grand canonical ensemble, the local and
global thermal phase structure of a charged asymptotically AdS black
hole with both GB and quartic field strength corrections were
thoroughly researched \cite{anninos}. In \cite{LYX} the phase
transition and critical phenamena of d-dimensional charged GB-AdS
black hole is analyzed. The consequence show that the phase
structure and critical temperature, critical electric charge,
critical electrostatic potential of the black hole all depend on the
cosmological constant $\Lambda $ and the dimension of spacetime. The
$(Q\sim U)$ critical conditions of d-dimensional charged GB-AdS
black hole agree with the $(P\sim V)$ones in van der Waals
vapor-liquid system.

Recently, many interests focus on the studies of critical behaviors
of AdS black holes \cite{RBM,RBM1,RBM2,dolan}by considering cosmological constant
as thermal pressure
\begin{equation}
\label{eq1} P=-\frac{1}{8\pi }\Lambda =\frac{3}{8\pi }\frac{1}{l^2},
\end{equation}
and corresponding conjugate thermal volume as
\begin{equation}
\label{eq2} V=\left( {\frac{\partial M}{\partial P}} \right)_{S,Q_i
,J_k } .
\end{equation}
The complete analog model of vapor-liquid system for black holes is
established in \cite{RBM}. In \cite{pope} the relation (\ref{eq1}) in higher
dimensional spherically symmetric AdS spacetime first proposed which
lays the foundation for the research of black holes thermodynamics.

Theoretically, if regarding the black holes in AdS spacetime as
thermodynamic systems the corresponding critical behaviors and phase
transition should exist. However, until now the statistical
explanation of black hole thermodynamics is still lack. Therefore it
is a meaningful work to discuss the relations of thermodynamic
properties for all kinds of black holes in AdS spacetime. This may
help to recognize further black hole entropy, temperature, heat
capacity and may help to improve the geometric theory of black holes
thermodynamics .

A scalar field called the dilaton appears in the low energy limit of
string theory. The presence of the dilaton field has important
consequences on the causal structure and the thermodynamic
properties of black holes. Thus much interest has been focused on
the study of the dilaton black holes in recent
years\cite{lemos1,gao,hendi1,ong,goldstein,CMC,BG,lizuka,WJL,WJL1,rocha,hendi,kachru,YCO,lala}.

In this paper we study the phase transition and critical behaviors
of $(n+1)$ dimensional charged topological dilaton AdS black hole.
Firstly we consider the cosmological constant as thermodynamic
pressure and the conjugate thermodynamic volume. We find that the
phase structure and critical phenomena are dependent on the static
electrostatic charge $Q$, dimension $n$ and dilaton field $\Phi $.
The critical exponents are the same as the ones in van der Waals
vapor-liquid system. Secondly we consider the pair of conjugate
parameters $(Q\sim U)$ as the thermodynamic variables and study the
phase transition and critical behaviors of $(n+1)$ dimensional
charged topological dilaton AdS black hole again. The results show
that the phase structure and critical phenomena are dependent on the
cosmological constant $\Lambda$, dimension $n$ and dilaton field
$\Phi $. The critical exponents are also the same as the ones in van
der Waals vapor-liquid system. Thus the two approaches are
equivalent because of the same phase diagrams and critical behavior.

The paper is arranged as follows: in the next section we simply
introduce the $(n+1)$-dimensional charge Dilaton AdS black hole. In
section 3  we will consider the parameters $(P\sim V)$ and $(Q\sim
U)$ respectively and discuss the phase structure and critical
phenomena of black holes. We will make some concluding remarks in
section 4. (we use the units $G_{n+1} =\hbar =k_B =c=1)$

\section{Charged Dilaton Black Holes in Anti-de Sitter Space}

The Einstein-Maxwell-Dilaton action in $(n+1)$-dimensional $(n\ge
3)$spacetime is [49,50]
\begin{equation}
\label{eq3} S=\frac{1}{16\pi }\int {d^{n+1}} x\sqrt {-g} \left(
{R-\frac{4}{n-1}(\nabla \Phi )^2-V(\Phi )-e^{-4\alpha \Phi
/(n-1)}F_{\mu \nu } F^{\mu \nu }} \right),
\end{equation}
where the dilaton potential is expressed in terms of the dilaton
field and its coupling to the cosmological constant:
\begin{equation}
\label{eq4} V^2\Phi =\frac{n-1}{8}\frac{\partial V}{\partial \Phi
}-\frac{\alpha }{2}e^{-4\alpha \Phi /(n-1)}F_{\lambda \eta }
F^{\lambda \eta },
\end{equation}
\begin{equation}
\label{eq5} \nabla _\mu \left( {e^{-4\alpha \Phi /(n-1)}F^{\mu \nu
}} \right)=0,
\end{equation}
where $R$ is the Ricci scalar curvature, $\Phi $ is the dilaton
field and $V(\Phi )$ is a potential for $\Phi $, $\alpha $ is a
constant determining the strength of coupling of the scalar and
electromagnetic field, $F_{\mu \nu } =\partial _\mu A_\nu -\partial
_\nu A_\mu $ is the electromagnetic field tensor and $A_\mu $ is the
electromagnetic potential. The topological black hole solutions take
the form [38,49,50]
\begin{equation}
\label{eq6} ds^2=-f(r)dt^2+\frac{dr^2}{f(r)}+r^2R^2(r)d\Omega
_{k,n-1}^2 ,
\end{equation}
where
\[
f(r)=-\frac{k(n-2)(\alpha ^2+1)^2b^{-2\gamma }r^{2\gamma }}{(\alpha
^2-1)(\alpha ^2+n-2)}-\frac{m}{r^{(n-1)(1-\gamma
)-1}}+\frac{2q^2(\alpha ^2+1)^2b^{-2(n-2)\gamma }}{(n-1)(\alpha
^2+n-2)}r^{2(n-2)(\gamma -1)}
\]
\begin{equation}
\label{eq7} -\frac{n(\alpha ^2+1)^2b^{2\gamma }}{l^2(\alpha
^2-n)}r^{2(1-\gamma )},
\end{equation}
\begin{equation}
\label{eq8} R(r)=e^{2\alpha \Phi /(n-1)}, \quad \Phi
(r)=\frac{(n-1)\alpha }{2(1+\alpha ^2)}\ln \left( {\frac{b}{r}}
\right),
\end{equation}
with $\gamma =\alpha ^2/(\alpha ^2+1)$. The cosmological constant is
related to spacetime dimension $n$ by
\begin{equation}
\label{eq9} \Lambda =-\frac{n(n-1)}{2l^2},
\end{equation}
where $b$ is an arbitrary constant and $l$ denotes the AdS length
scale . In the above expression, $m$ appears as an integration
constant and is related to the ADM (Arnowitt-Deser-Misnsr) mass of
the black hole. According to the definition of mass due to Abbott
and Deser \cite{deser,olea}, the mass of the solution (\ref{eq7}) is \cite{sheykhi}
\begin{equation}
\label{eq10} M=\frac{b^{(n-1)\gamma }(n-1)\omega _{n-1} }{16\pi
(\alpha ^2+1)}m.
\end{equation}
the electric charge is
\begin{equation}
\label{eq11} Q=\frac{q\omega _{n-1} }{4\pi },
\end{equation}
where $\omega _{n-1} $ represents the volume of constant curvature
hypersurface described by $d\Omega _{k,n-1}^2 $. The Hawking
temperature of the topological black hole on the outer horizon $r_+
$ can be calculated using the relation
\begin{equation}
\label{eq12} T=\frac{\kappa }{2\pi }=\frac{f'(r_+ )}{4\pi },
\end{equation}
where $\kappa $ is the surface gravity. It can easily show that
\[
T=-\frac{(\alpha ^2+1)}{2\pi (n-1)}\left(
{\frac{k(n-2)(n-1)b^{-2\gamma }}{2(\alpha ^2-1)}r_+^{2\gamma -1}
+\Lambda b^{2\gamma }r_+^{1-2\gamma } +q^2b^{-2(n-2)\gamma
}r_+^{(2n-3)(\gamma -1)-\gamma } } \right)
\]
\[
=-\frac{k(n-2)(\alpha ^2+1)b^{-2\gamma }}{2\pi (\alpha
^2+n-2)}r_+^{2\gamma -1} +\frac{(n-\alpha ^2)m}{4\pi (\alpha
^2+1)}r_+^{(n-1)(\gamma -1)}
\]
\begin{equation}
\label{eq13} -\frac{q^2(\alpha ^2+1)b^{-2(n-2)\gamma }}{\pi (\alpha
^2+n-2)}r_+^{(2n-3(\gamma -1)-\gamma } .
\end{equation}
Topological black hole entropy
\begin{equation}
\label{eq14} S=\frac{b^{(n-1)\gamma }\omega _{n-1}
r_+^{(n-1)(1-\gamma )} }{4}.
\end{equation}
The electric potential
\begin{equation}
\label{eq15} U =\frac{qb^{(3-n)\gamma }}{r_+^\lambda \lambda },
\end{equation}
where $\lambda =(n-3)(1-\gamma )+1$. From $f(r_+ )=0$ and
(\ref{eq10}), we obtain
\[
M=\frac{q^2b^{\gamma (3-n)}(\alpha ^2+1)\omega _{n-1} }{8\pi (\alpha
^2+n-2)}r_+^{-\lambda } -\frac{n(n-1)(\alpha ^2+1)b^{\gamma
(n+1)}\omega _{n-1} }{16\pi l^2(\alpha ^2-n)}r_+^{n-\gamma (n+1)}
\]
\begin{equation}
\label{eq16} -\frac{k(n-2)(n-1)(\alpha ^2+1)b^{\gamma (n-3)}\omega
_{n-1} }{16\pi (\alpha ^2-1)(\alpha ^2+n-2)}r_+^\lambda .
\end{equation}
One may then regard the parameters $S$, $Q$ and $P$ as a complete
set of extensive parameters for the mass $M(S,Q,P)$ and define the
intensive parameters conjugate to $S$, $Q$ and $P$. These quantities
are the temperature, the electric potential and volume.
\begin{equation}
\label{eq17} T=\left( {\frac{\partial M}{\partial S}} \right)_{Q,P}
, \quad U=\left( {\frac{\partial M}{\partial Q}} \right)_{S,P} ,
\quad V=\left( {\frac{\partial M}{\partial P}} \right)_{Q,S} .
\end{equation}
where
\begin{equation}
\label{eq18} P=\frac{n(n-1)}{16\pi l^2}, \quad V=-\frac{(\alpha
^2+1)b^{\gamma (n+1)}\omega _{n-1} }{(\alpha ^2-n)}r_+^{n-\gamma
(n+1)} .
\end{equation}
It is a matter of straightforward calculation to show that the
quantities calculated by Eq. (\ref{eq18}) for the temperature, and
the electric potential coincide with Eqs. (\ref{eq13}) and
(\ref{eq15}). Thus, the thermodynamics quantities satisfy the first
law of thermodynamics
\begin{equation}
\label{eq19} dM=TdS+UdQ+Vdp.
\end{equation}
The thermodynamic quantities above, energy $M$, entropy $S$,
temperature $T$, volume $V$, pressure $P$, electrostatic potential
$U$ and electric charge $Q$ satisfy Smarr formula:
\begin{equation}
\label{eq20} M=\frac{(n-1)(1-\gamma )}{\lambda
}TS+UQ-\frac{n-\lambda }{\lambda }VP.
\end{equation}
In what follows we concentrate on analyzing the phase transition of
the $(n+1)$dimensional charged topological dilaton AdS black hole
system in the extended phase space while we treat the black hole
charge $Q$ as a fixed external parameter, or the cosmological
constant is an invariant parameter, not a thermodynamic variable. We
shall find that an even more remarkable coincidence with the Van der
Waals fluid is realized in this case.

\section{Critical behaviour}

\subsection{$Q$ {\bf is an invariant parameter}}

For a fixed charge $q$, Eq. (\ref{eq13}) translates into the
equation of state for a charged topological dilaton black hole,
$P=P(V,T)$
\[
P=\frac{T(n-1)}{4(\alpha ^2+1)}b^{-2\gamma }r_+^{2\gamma -1}
+\frac{k(n-1)(n-2)}{16\pi (\alpha ^2-1)}b^{-4\gamma }r_+^{2(2\gamma
-1)} +\frac{Q^2b^{2(1-n)\gamma }2\pi }{\omega _{n-1}^2
}r_+^{2(n-1)(\gamma -1)}
\]
\begin{equation}
\label{eq21} r_+ =\left( {\frac{V(n-\alpha ^2)}{(\alpha ^2+1)\omega
_{n-1} b^{\gamma (n+1)}}} \right)^{1/(n-\gamma (n+1))}.
\end{equation}
Where $P$ and $V$ are given by (\ref{eq18}), $V$ is the
thermodynamic volume, given in terms of the event horizon radius
$r_+ $, $T$ is the black hole temperature, and $Q$ its charge. The
Van der Waals equation
\begin{equation}
\label{eq22} \left( {P+\frac{a}{v^2}} \right)(v-\tilde {b})=kT,
\end{equation}
Here, $v=V/N$ is the specific volume of the fluid, $P$ its pressure,
$T$ its temperature, and $k$ is the Boltzmann constant.

Comparing with the Van der Waals equation, (\ref{eq22}), we conclude
that we should identify the specific volume $v$ of the fluid with
the horizon radius of the black hole as
\begin{equation}
\label{eq23} v=\frac{4(\alpha ^2+1)b^{2\gamma
}}{(n-1)}r_+^{1-2\gamma } .
\end{equation}
In $(n+1)$ dimensions, the equation of (\ref{eq21}) reads
\[
P=\frac{T}{v} +\frac{k(n-2)(\alpha ^2+1)^2}{\pi (n-1)(\alpha
^2-1)v^2} +\frac{Q^2b^{2(1-n)\gamma }2\pi }{\omega _{n-1}^2 }\left(
{\frac{v(n-1)}{4(\alpha ^2+1)b^{2\gamma }}}
\right)^{\textstyle{{2(n-1)(\gamma -1)} \over {1-2\gamma }}}
\]
\begin{equation}
\label{eq24} =\frac{T}{v} -\frac{A}{v^2}
+\frac{B}{v^{2(n-1)(1-\gamma )/(1-2\gamma )}},
\end{equation}
where
\begin{equation}
\label{eq25} A=\frac{k(n-2)(\alpha ^2+1)^2}{\pi (n-1)(1-\alpha ^2)},
\quad B=\frac{Q^2b^{2(1-n)\gamma }2\pi }{\omega _{n-1}^2 }\left(
{\frac{4(\alpha ^2+1)b^{2\gamma }}{(n-1)}} \right)^{2(n-1)(1-\gamma
)/(1-2\gamma )}.
\end{equation}
Critical points occur at points of inflection in the $P-V$ diagram,
where
\begin{equation}
\label{eq26} \frac{\partial P}{\partial v}=0, \quad \frac{\partial
^2P}{\partial v^2}=0,
\end{equation}
Substituting (\ref{eq24}) into (\ref{eq26}) we can derive the
critical volume, temperature and pressure:
\[
v_c^{x-2} =\frac{(x-1)xB}{2A}, \quad T_c =\frac{2A}{v_c
}-\frac{xB}{v_c^{x-1} } =\frac{2A(x-2)}{(x-1)}\left(
{\frac{2A}{(x-1)xB}} \right)^{1/(x-2)},
\]
\begin{equation}
\label{eq27} P_c =\frac{A}{v_c^2 }-\frac{B}{v_c^x }(x-1)
=\frac{A(x-2)}{x}\left( {\frac{2A}{(x-1)xB}} \right)^{2/(x-2)},
\end{equation}
where
\begin{equation}
\label{eq28} x=\frac{2(n-1)(1-\gamma )}{(1-2\gamma )}.
\end{equation}
Van der Waals equation critical temperature, volume and pressure
are:
\[
v_c^{x-2} =\frac{(x-1)xB}{2A}, \quad T_c =\frac{2A}{v_c
}-\frac{xB}{v_c^{x-1} } =\frac{2A(x-2)}{(x-1)}\left(
{\frac{2A}{(x-1)xB}} \right)^{1/(x-2)},
\]
\begin{equation}
\label{eq29} T_c =\frac{8a}{27\tilde {b}}, \quad v_c =3\tilde {b},
\quad P_c =\frac{a}{27\tilde {b}^2}.
\end{equation}
From (\ref{eq27}), $(n+1)$-dimensional charged topological dilaton
black hole correspond to
\begin{equation}
\label{eq30} \tilde {b}=\frac{x}{4(x-1)}\left( {\frac{(x-1)xB}{2A}}
\right)^{1/(x-2)}, \quad a=\frac{27Ax(x-2)}{16(x-1)^2}, \quad \rho
_c =\frac{P_c v_c }{T_c }=\frac{x-1}{2x}.
\end{equation}
Therefore, the critical temperature, volume and pressure of
$(n+1)$-dimensional charged topological dilaton black hole are
\begin{equation}
\label{eq31} v_c =\frac{4(x-1)}{x}\tilde {b}, \quad T_c
=\frac{8a}{27\tilde {b}}, \quad P_c =\frac{a}{27\tilde {b}^2}.
\end{equation}

\begin{figure}\label{pv}
\center{
\subfigure[($n=3, \alpha =0$)]{
\label{1-a}
\includegraphics[scale=0.33]{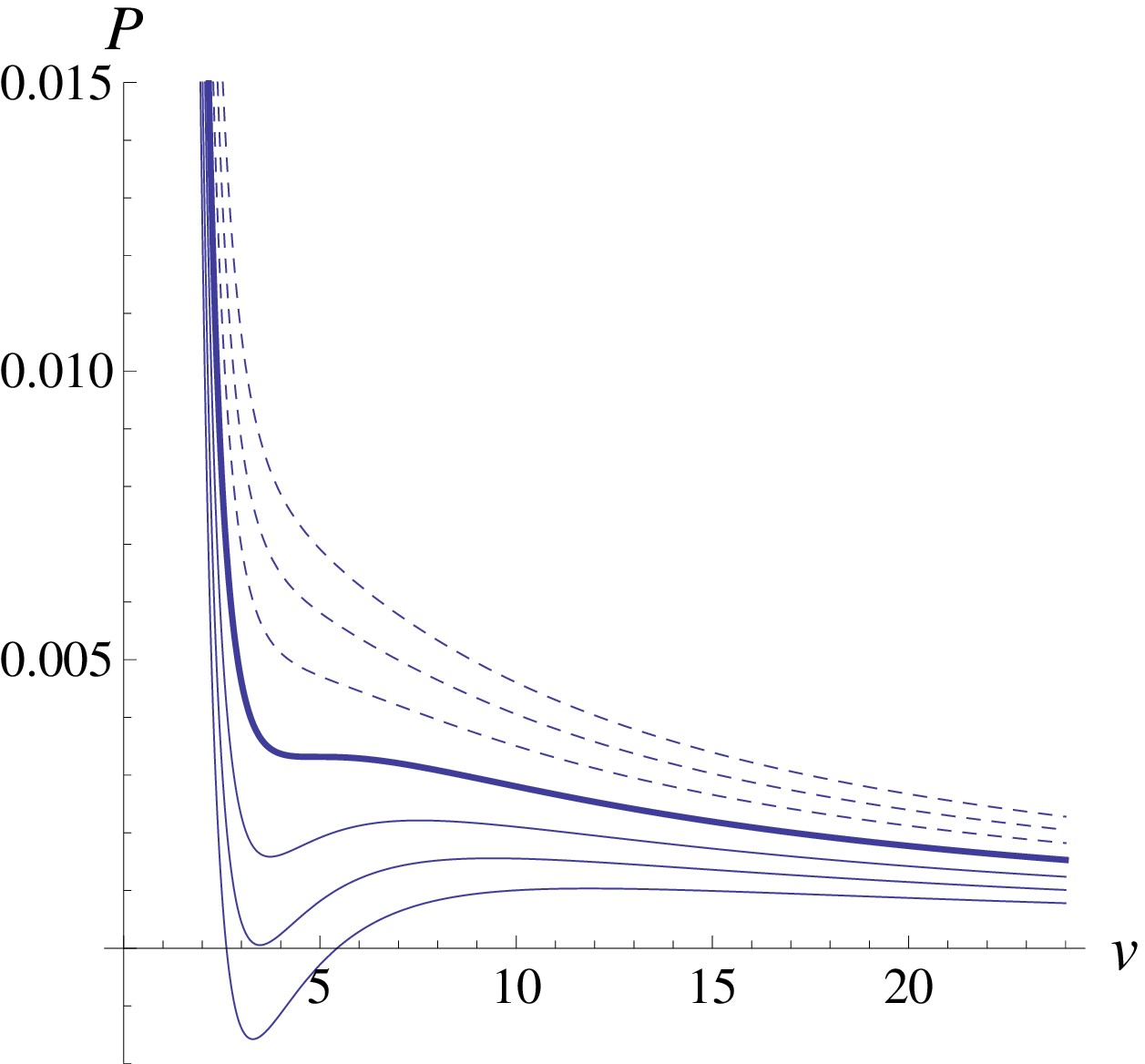}\hspace{0.5cm}}
\subfigure[($n=3, \alpha =0.6$)]{
\label{1-b}
\includegraphics[scale=0.33]{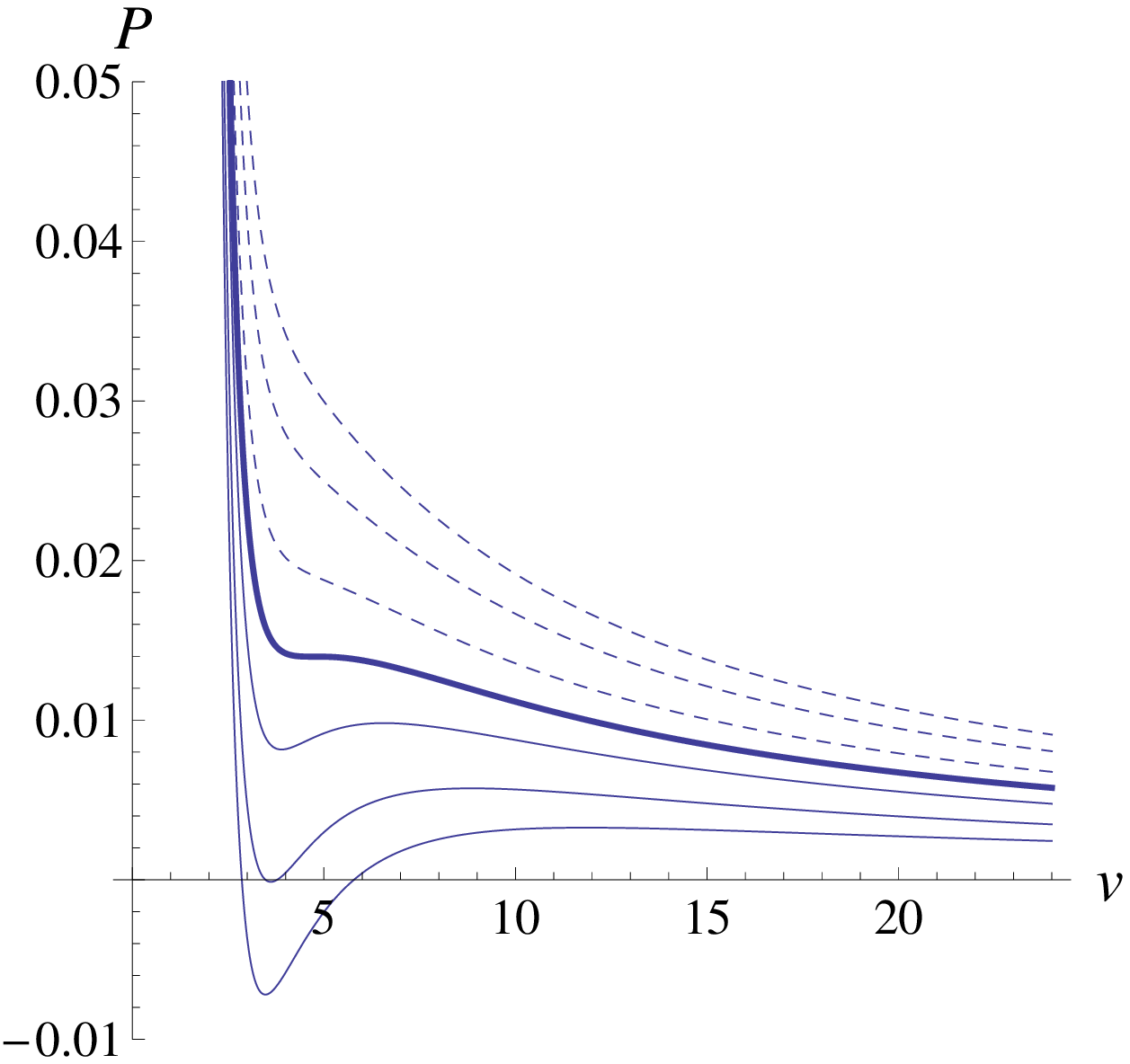}\hspace{0.5cm}}
\subfigure[($n=10, \alpha =0.6$)]{
\label{1-c}
\includegraphics[scale=0.33]{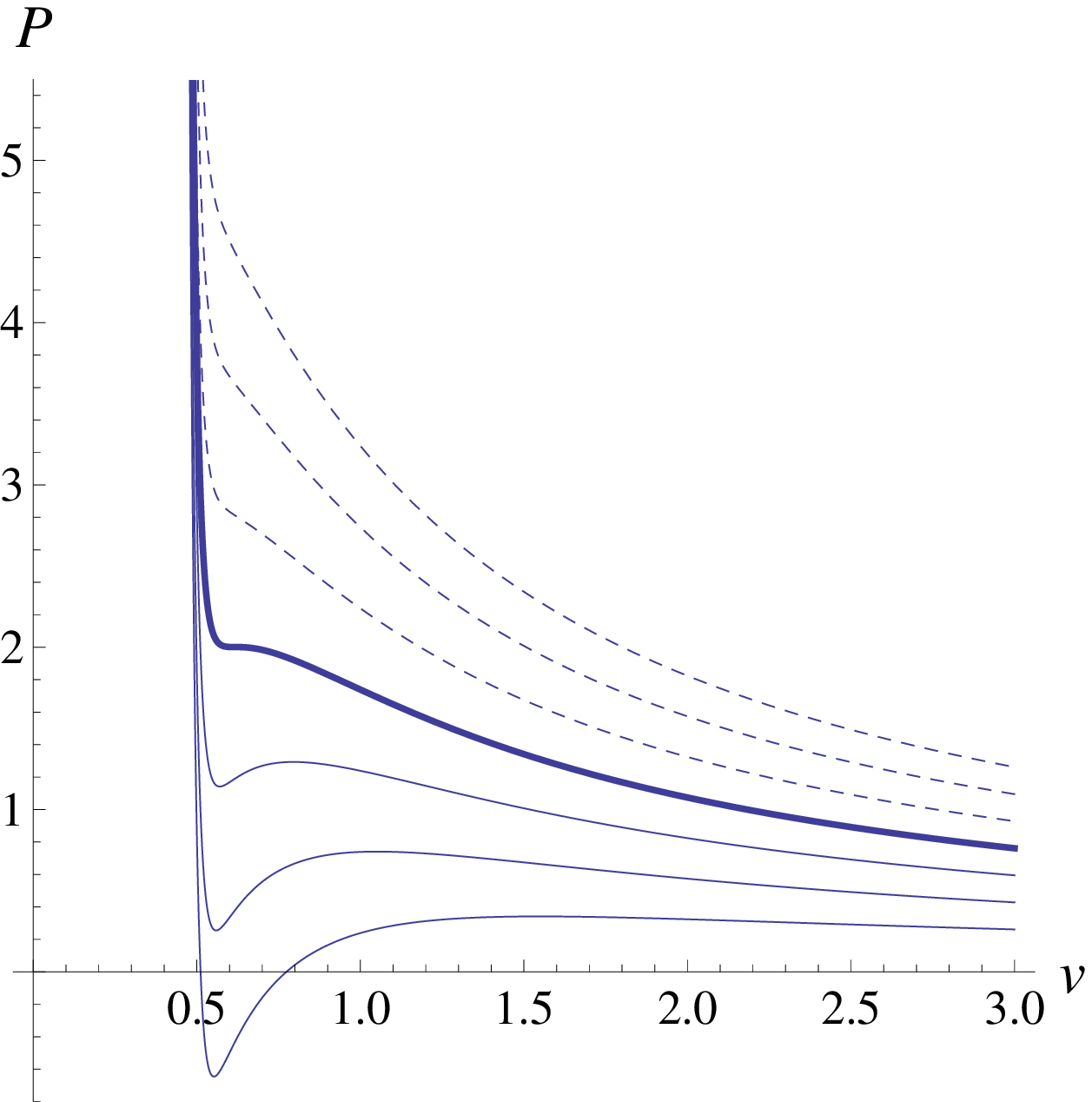}\hspace{0.5cm}}
\caption{$p-v$ diagram of $(n+1)$-dimensional charge Dilaton AdS
black hole in ($n=3, \alpha =0$), ($n=3, \alpha =0.6$) and ($n=10, \alpha =0.6$) respectively. The temperature
of isotherms decreases from top to bottom. The three upper dashed
lines correspond to the ``ideal gas'' one-phase behaviour for $T>T_c
$, the critical isotherm $T=T_c $ is denoted by the thick solid
line, lower (red) solid lines correspond to two-phase state
occurring for $T<T_c $. We have set $Q=1,b=1,k=1$. The behaviour for
$n>3$ and $\alpha >0$ is qualitatively.}}
\end{figure}

\begin{figure}
\center{
\includegraphics[scale=0.45]{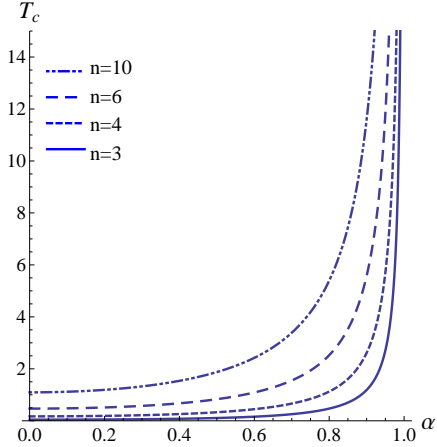}\hspace{1cm}
\caption{The $T_c -\alpha $ diagram shows the influence of dilaton
field $\alpha $ on the critical temperature with different sapcetime
dimensions.}}
\end{figure}

In Fig.1, (a) and (b) show the influence of dilaton field
$\alpha $ on the isothermal curves with the same spacetime
dimension. (b) and (c) represent the influence of
spacetime dimension $n$ on the isothermal curves with the same
dilaton field.

To calculate the critical exponent $\alpha $ we consider the entropy
$S$, (\ref{eq14}), as a function of $T$ and $V$ . Using (\ref{eq21})
we have
\begin{equation}
\label{eq32} S=S(T,V) =\frac{b^{-\gamma (n-1)/(n-\gamma (n+1))}}{4}
\omega _{n-1}^{(1-2\gamma )/(n-\gamma (n+1))} \left(
{\frac{V(n-\alpha ^2)}{(\alpha ^2+1)}} \right)^{(n-1)(1-\gamma
)/(n-\gamma (n+1))}.
\end{equation}
Since this is independent of $T$, we have $C_V =T\left(
{\frac{\partial S}{\partial T}} \right)_V =0$ and hence $\tilde
{\alpha }=0$. Defining specific variables
\begin{equation}
\label{eq33} p=\frac{P}{P_c }, \quad v=\frac{v}{v_c }, \quad \tau
=\frac{T}{T_c }.
\end{equation}
Thus Eq.(\ref{eq24}) turns into
\begin{equation}
\label{eq34} p=\frac{\tau
}{v}\frac{2x}{x-1}-\frac{1}{v^2}\frac{x}{x-2}
+\frac{1}{v^x}\frac{2}{(x-1)(x-2)},
\end{equation}
One can write Eq.(\ref{eq34}) in the form of van der Waals
\begin{equation}
\label{eq35} (x-1)v\left( {p+\frac{1}{v^2}\frac{x}{x-2}}
\right)-\frac{2}{v^{x-1}(x-2)}=2x\tau .
\end{equation}
In Eq.(\ref{eq34}) there exist no constants which depend on the
properties of matter, but there is the quantity $x$ which is dependent on  $n$ and $\gamma $.
 Therefore, when $n$ and $\gamma$ are the same, the
equation can be simplified to
\begin{equation}
\label{eq36} p=\frac{\tau }{\rho _c v}+f(v),
\end{equation}
where $\rho _c $ stand for the critical ratio, which is given by
(\ref{eq30}).
\begin{equation}
\label{eq37} f(v)= -\frac{1}{v^2}\frac{x}{x-2}
+\frac{1}{v^x}\frac{2}{(x-1)(x-2)}
\end{equation}
where $\rho _c $ stands for the critical ratio. Expanding this
equation near the critical point
\begin{equation}
\label{eq38} \tau =t+1, \quad v=\left( {\omega +1} \right)^{1/q},
\end{equation}
where $\tilde {q}>0$, and using the fact that from the definition of
the critical point we have
\begin{equation}
\label{eq39} \frac{1}{\rho _c }+f(1)=1, \quad \rho _c f'(1)=1, \quad
\rho _c f''(1)=-2,
\end{equation}
And so obtain
\begin{equation}
\label{eq40} p=1+\frac{t}{\rho _c }-\frac{t\omega }{\tilde {q}\rho
_c }-C\omega ^3+O(t\omega ^2,\omega ^4),
\end{equation}
where $C=\frac{1}{\tilde {q}^3}\left( {\frac{1}{\rho _c
}-\frac{f^{(3)}(1)}{6}} \right)$.Differentiating the series for a
fixed $t<0$ we get
\begin{equation}
\label{eq41} dP=-P_c \left( {\frac{t}{\tilde {q}\rho _c }+3C\omega
^2} \right)d\omega .
\end{equation}
Employing Maxwell's equal area law, see, e.g., \cite{LYX,RBM,RBM2},
while denoting $\omega _g$ and $\omega _l$ the `volume' of small and
large black holes, we get the following two equations:
\[
p=1+\frac{t}{\rho _c }-\frac{t\omega _l }{\tilde {q}\rho _c
}-C\omega _l^3 =1+\frac{t}{\rho _c }-\frac{t\omega _g }{\tilde
{q}\rho _c }-C\omega _g^3 ,
\]
\begin{equation}
\label{eq42} 0=\int\limits_{\omega _l }^{\omega _g } {\omega dP=}
\int\limits_{\omega _l }^{\omega _g } {\omega \left(
{\frac{t}{\tilde {q}\rho _c }+3C\omega ^2} \right)d\omega } .
\end{equation}
The unique non-trivial solution is
\begin{equation}
\label{eq43} \omega _g =-\omega _l = \sqrt {-\frac{2t}{3C\tilde
{q}\rho _c }} \propto (-t)^{1.2},
\end{equation}
which implies that the degree of the coexistence curve
\begin{equation}
\label{eq44} \beta =1/2.
\end{equation}
To calculate the exponent $\tilde {\gamma }$, we use again
(\ref{eq42}), to get
\begin{equation}
\label{eq45} \kappa _T =-\frac{1}{V}\left( {\frac{\partial
V}{\partial P}} \right)_T \propto \frac{1}{P_c }\frac{1}{\left(
{t/(\tilde {q}\rho _c +3C\omega ^2} \right)},
\end{equation}
The set $\omega =0$, we get
\begin{equation}
\label{eq46} \kappa _T \propto \frac{1}{P_c }\frac{\tilde {q}\rho _c
}{t}.
\end{equation}
Thus the isothermal compressibility exponent
\begin{equation}
\label{eq47} \tilde {\gamma }=1.
\end{equation}
Finally, the `shape of the critical isotherm' $t=0$ is given by
(\ref{eq43}), i.e.,
\begin{equation}
\label{eq48} p-1=-C\omega ^3,
\end{equation}
Therefore the critical exponent
\begin{equation}
\label{eq49} \delta =3.
\end{equation}

According to Eq.(\ref{eq31}), when the electric charge of black
holes is invariant, the equation of state of the charged topological
dilaton AdS black hole in any dimension can be expressed as the form
of Van der Waals equation. The critical exponents are the same as
the ones for Van der Waals fluid.

In particular, the law of corresponding states, (\ref{eq34}),
takes the form (\ref{eq36}). Taking $\tilde {q}=\frac{n-\gamma
(n+1)}{1-2\gamma }$ (in which cases $\omega =\frac{V}{V_c }-1)$. We
obtain the expansion (\ref{eq41}) with
\begin{equation}
\label{eq50} C=\frac{x(1-2\gamma )^3}{3[n-\gamma (n+1)]^3},
\end{equation}
and so the discussion above applies.

\subsection{$l$ {\bf is an invariant parameter}}

In the case of $l$ invariant, substituting Eq.(\ref{eq15}) into
(\ref{eq13}), one can derive
\[
T=\frac{(\alpha ^2+1)}{4\pi }\frac{k(n-2)b^{-2\gamma }}{(1-\alpha
^2)}\left( {\frac{qb^{(3-n)\gamma }}{\lambda U}} \right)^{(2\gamma
-1)/\lambda } +\frac{n(\alpha ^2+1)b^{2\gamma }}{4\pi l^2}\left(
{\frac{qb^{(3-n)\gamma }}{\lambda U}} \right)^{(1-2\gamma )/\lambda
}
\]
\[
-\frac{(\alpha ^2+1)}{2\pi (n-1)}q^2b^{-2(n-2)\gamma }\left(
{\frac{qb^{(3-n)\gamma }}{\lambda U}} \right)^{((2n-3)(\gamma
-1)-\gamma )/\lambda }
\]
\begin{equation}
\label{eq51} =\tilde {A}\left( {\frac{qb^{(3-n)\gamma }}{\lambda U}}
\right)^{(2\gamma -1)/\lambda } +\tilde {B}\left(
{\frac{qb^{(3-n)\gamma }}{\lambda U}} \right)^{(1-2\gamma )/\lambda
} -\tilde {C}q^2\left( {\frac{qb^{(3-n)\gamma }}{\lambda U}}
\right)^{((2n-3)(\gamma -1)-\gamma )/\lambda },
\end{equation}
where
\begin{equation}
\label{eq52} \tilde {A}=\frac{(\alpha ^2+1)}{4\pi
}\frac{k(n-2)b^{-2\gamma }}{(1-\alpha ^2)},\; \tilde
{B}=\frac{n(\alpha ^2+1)b^{2\gamma }}{4\pi l^2}, \;\tilde
{C}=\frac{(\alpha ^2+1)}{2\pi (n-1)}b^{-2(n-2)\gamma }.
\end{equation}
The critical points should satisfy the conditions£º
\begin{equation}
\label{eq53} \left( {\frac{\partial q}{\partial U}} \right)_T
=\left( {\frac{\partial ^2q}{\partial U^2}} \right)_T =0.
\end{equation}
Substituting Eq.(\ref{eq51}) into Eq.(\ref{eq53}), one can obtain
the critical electric charge, critical electrostatic potential and
critical temperature:
\[
q_c^2 =-\frac{\tilde {A}'+\tilde {B}'D}{\tilde {C}'D^{[n\gamma
-3\gamma -n+2]/(1-2\gamma )}}=\frac{k(n-2)}{2(x-1)}b^{2(n-3)\gamma
}D^{\lambda /(1-2\gamma )}=E
\]
\[
\lambda U_c =E^{1/2}b^{(3-n)\gamma }D^{-\lambda /(2-4\gamma )},
\]
\begin{equation}
\label{eq54} T_c =\tilde {A}D^{-1/2} +\tilde {B}D^{1/2} +\tilde
{C}\frac{\tilde {A}'+\tilde {B}'D}{\tilde {C}'}D^{-1/2}
=\frac{k(n-2)b^{-2\gamma }(x-2)}{2\pi (1-2\gamma )(x-1)}D^{-1/2},
\end{equation}
where
\[
\tilde {A}'=\tilde {A}(1-2\gamma ), \quad \tilde {B}'=\tilde {B}(2\gamma
-1), \quad \tilde {C}'=\tilde {C}[(2n-3)(\gamma -1)-\gamma ], \quad
D=\frac{k(n-2)b^{-4\gamma }l^2}{2n(n-1)}(x-2).
\]

\begin{figure}\label{qu}
\center{ \subfigure[($n=3, \alpha =0$)]{ \label{3-a}
\includegraphics[scale=0.33]{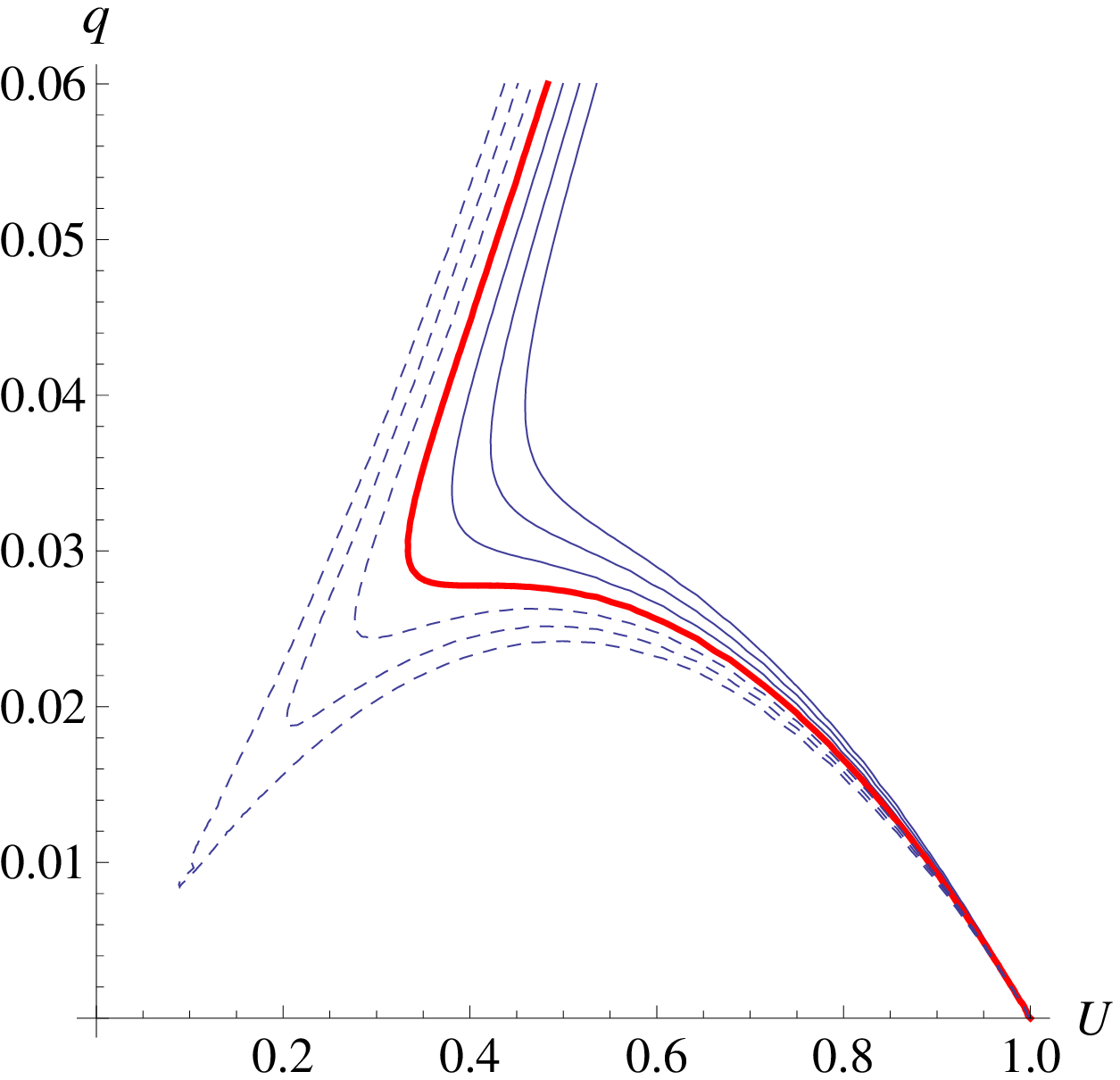}\hspace{0.5cm}}
\subfigure[($n=3, \alpha =0.6$)]{ \label{3-b}
\includegraphics[scale=0.33]{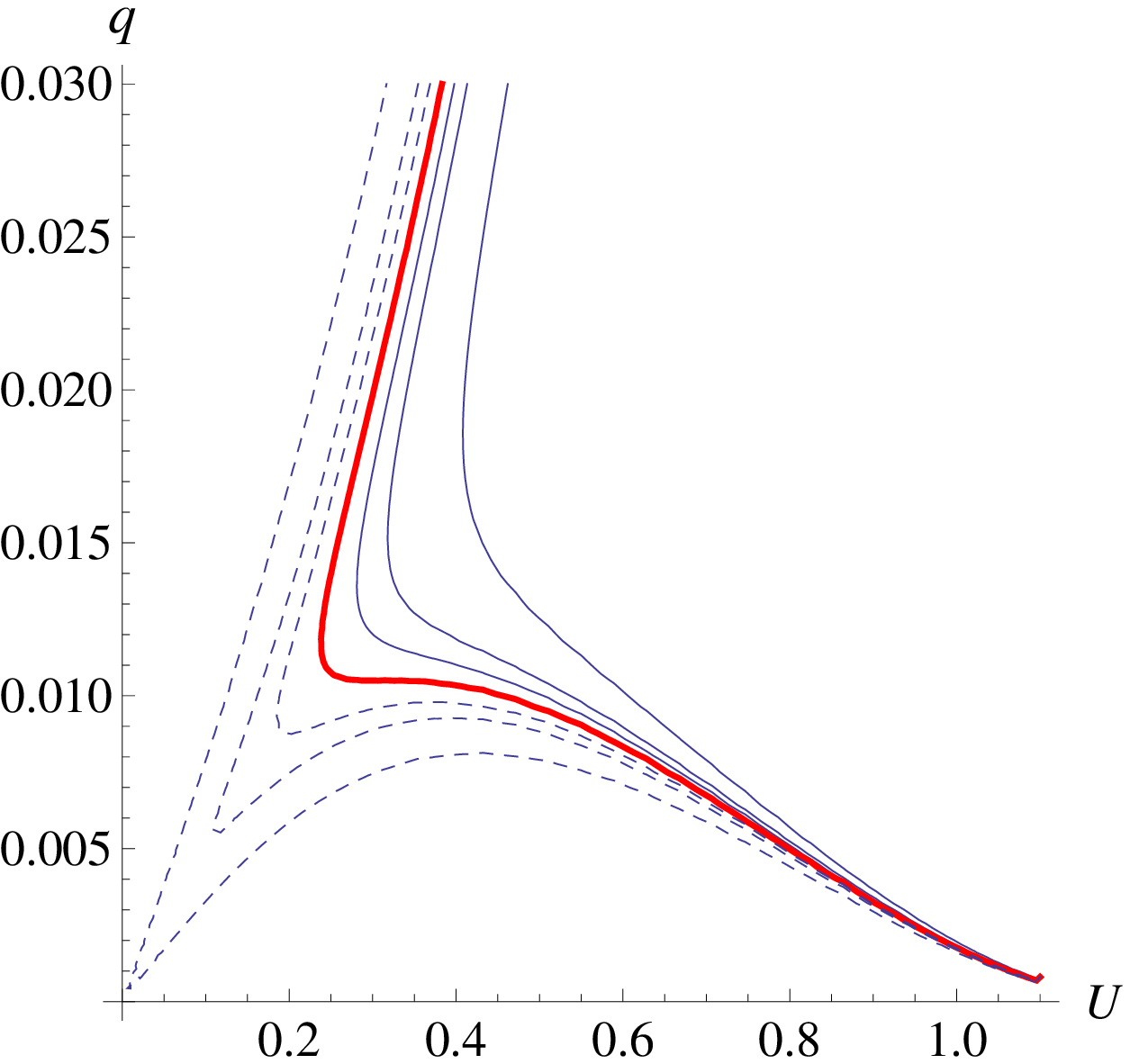}\hspace{0.5cm}}
\subfigure[($n=10, \alpha =0.6$)]{ \label{3-c}
\includegraphics[scale=0.33]{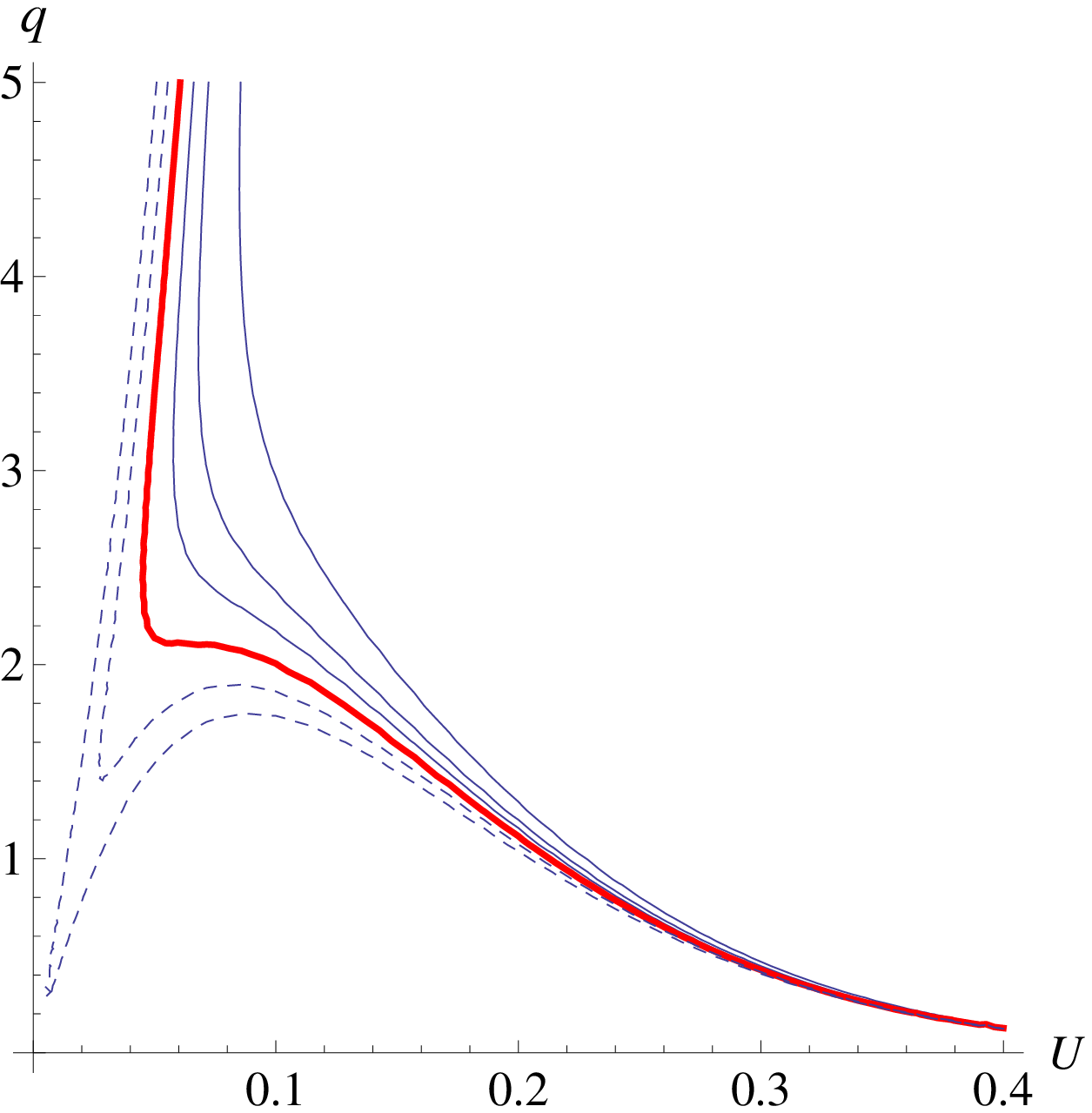}\hspace{0.5cm}}
\caption{$q-U$ diagram of $(n+1)$-dimensional charge Dilaton AdS
black hole in ($n=3, \alpha =0$), ($n=3, \alpha =0.6$) and ($n=10,
\alpha =0.6$) respectively. The temperature of isotherms decreases
from top to bottom. The three lower dashed lines correspond to the
``ideal gas'' one-phase behaviour for $T>T_c$, the critical isotherm
$T=T_c $ is denoted by the thick solid line, upper (red) solid lines
correspond to two-phase state occurring for $T<T_c $. We have set
$b=1,k=1$ and the cosmological constant is given by  Eqs.(3.5), (3.7).  The behaviour for $n>3$ and
$\alpha>0$ is qualitatively.
}}
\end{figure}

In Fig.3, (a) and (b) show the influence of dilaton field
$\alpha $ on the isothermal curves with the same spacetime
dimension. (b) and (c) represent the influence of
spacetime dimension $n$ on the isothermal curves with the same
dilaton field.

Expanding ground a critical point
$\tau =\frac{\beta }{\beta _c }$£¬ $u=\frac{U}{U_c }$£¬ $\vartheta
=\frac{q}{q_c }$. Eq.(\ref{eq51}) can be expressed as
\begin{equation}
\label{eq55} \frac{1}{\tau \beta _c }=\tilde {A}D^{-1/2}\left(
{\frac{\vartheta }{u}} \right)^{(2\gamma -1)/\lambda } +\tilde
{B}D^{1/2}\left( {\frac{\vartheta }{u}} \right)^{(1-2\gamma
)/\lambda } +\tilde {C}\frac{\tilde {A}'+\tilde {B}'D}{\tilde
{C}'}\vartheta ^2\left( {\frac{\vartheta }{u}}
\right)^{((2n-3)(\gamma -1)-\gamma )/\lambda },
\end{equation}
Near the critical points, defining $t=\tau -1$,\;$\omega =u-1$ and
substituting them into the above equation, we can obtain
\begin{equation}
\label{eq56} \vartheta =\sum\limits_{m,n=0} {a_{mn} } t^m\omega ^n,
\end{equation}
where $a_{mn} =\frac{1}{m!n!}\left. {\frac{\partial ^{(m+n)}\vartheta
}{\partial t^m\partial \omega ^n}} \right|_{\begin{array}{l}
 t=0 \\
 \omega =0 \\
 \end{array}}$. According to Eq.(\ref{eq51}) and (\ref{eq53}), $a_{00} =1$,\;$a_{01}
=0$,\;$a_{02} =0$. This result holds for any values of $k$, $l$and
$\gamma $. In the neighborhood of the critical points, we have
(\ref{eq56}). The values of $\omega $ on either side of the
coexistence curve can be found from the conditions that along the
isotherm, Employing Maxwell's equal area law, see, e.g., [33,35],
while denoting $\omega _g $ and $\omega _l $ the `volume' of small
and large black holes, we get the following two equations:
\begin{equation}
\label{eq57} \vartheta (\omega _g )=\vartheta (\omega _l ),
\end{equation}
\begin{equation}
\label{eq58} 0=\int\limits_{\omega _l }^{\omega _g } {(\omega +1)d}
\vartheta .
\end{equation}
From the first condition (\ref{eq57}), we derive
\begin{equation}
\label{eq59} a_{11} t(\tilde {\omega }_g +\tilde {\omega }_l
)+a_{21} t^2(\tilde {\omega }_g +\tilde {\omega }_l )+a_{12}
t(\tilde {\omega }_g^2 -\tilde {\omega }_l^2 )+a_{03} (\tilde
{\omega }_g^3 +\tilde {\omega }_l^3 )+o(t\omega ^2,\omega ^4)=0,
\end{equation}
where we have denoted $\tilde {\omega }_l =-\omega _l $ and $\tilde
{\omega }_g =\omega _g $ for different phases. The second condition
(\ref{eq58}) reduces to
\[
a_{11} t(\tilde {\omega }_g +\tilde {\omega }_l )+a_{21} t^2(\tilde
{\omega }_g +\tilde {\omega }_l )+\frac{1}{2}(a_{11} +2a_{12}
)t(\tilde {\omega }_g^2 -\tilde {\omega }_l^2 )
\]
\begin{equation}
\label{eq60} +a_{03} (\tilde {\omega }_g^3 +\tilde {\omega }_l^3
)+o(t\omega ^2,\omega ^4)=0.
\end{equation}
The unique non-trivial solution is
\begin{equation}
\label{eq61} \tilde {\omega }_g^2 =\tilde {\omega }_l^2
=-\frac{1}{a_{03} }(a_{11} t+a_{21} t^2).
\end{equation}
yielding
\begin{equation}
\label{eq62} \tilde {\omega }_g =\tilde {\omega }_l \sim \sqrt
{-\frac{a_{11} }{a_{03} }t} \sim t^{1/2} \quad \Rightarrow \quad
\beta =\frac{1}{2}.
\end{equation}
To calculate the exponent$\tilde {\gamma }$, we use again
(\ref{eq56}), to get
\begin{equation}
\label{eq63} \kappa _T =-\frac{1}{V}\left. {\frac{\partial
V}{\partial P}} \right|_T \propto \frac{1}{P_c }\frac{1}{a_{11} t}
\quad \Rightarrow \quad \tilde {\gamma }=1.
\end{equation}
Let $t=0$, (\ref{eq60}) will be of the form
\begin{equation}
\label{eq64} \vartheta =a_{03} \omega ^3+o(\omega ^4),
\end{equation}
Thus we get the degree of the critical isotherm
\begin{equation}
\label{eq65} \delta =3.
\end{equation}
The heat capacity at fixed $U$ is
\begin{equation}
\label{eq66} C_U =T\left( {\frac{\partial S}{\partial T}} \right)_U
=T\left( {\frac{\partial S}{\partial q}} \right)_U \left(
{\frac{\partial T}{\partial q}} \right)_U^{-1} ,
\end{equation}
From Eq.(\ref{eq14}) and Eq.(\ref{eq51})
\[
\left( {\frac{\partial S}{\partial q}} \right)_U
=\frac{b^{(n-1)\gamma }\omega _{n-1} }{4}\frac{(n-1)(1-\gamma
)}{\lambda }\left( {\frac{qb^{(3-n)\gamma }}{\lambda U}}
\right)^{(n-1)(1-\gamma )/\lambda }\frac{1}{q},
\]
\[
\left( {\frac{\partial T}{\partial q}} \right)_U =-\tilde
{A}\frac{(1-2\gamma )}{\lambda q}\left( {\frac{qb^{(3-n)\gamma
}}{\lambda U}} \right)^{(2\gamma -1)/\lambda } +\tilde
{B}\frac{(1-2\gamma )}{\lambda q}\left( {\frac{qb^{(3-n)\gamma
}}{\lambda U}} \right)^{(1-2\gamma )/\lambda }
\]
\[
+\tilde {C}q\frac{(1-2\gamma )}{\lambda }\left(
{\frac{qb^{(3-n)\gamma }}{\lambda U}} \right)^{((2n-3)(\gamma
-1)-\gamma )/\lambda }
\]
\begin{equation}
\label{eq67} =\frac{k(n-2)(x-2)b^{-2\gamma }}{2\pi \lambda q_c
x(x-1)}D^{-1/2},
\end{equation}
When $n\ne 2$, and $x\ne 2$, Eq.(\ref{eq67}) is nonzero and $C_U $
is non-singular at the critical points. Thus the heat capacity
exponent
\begin{equation}
\label{eq68} \tilde {\alpha }=\tilde {\alpha }'=0.
\end{equation}
When the cosmological constant is invariant we conclude that the
thermodynamic exponents associated with the charged topological
dilaton AdS black holes in any dimension $n\ge 3$ coincide with
those of the Van der Waals fluid.

\section{Discussion and Conclusions}
In Sec.3 we discussed the phase structure and critical
phenomena of charged topological dilaton AdS black holes in the case
of $k=1$ and the cases of electric charge $Q$ and the cosmological
constant $l$ are invariant respectively. We obtain two pairs of
critical temperature and critical pressure (critical electric
charge) and critical volume(critical electric potential), which are
represented by Eq.(\ref{eq27}) and (\ref{eq54}). Below we will
analyze the relations between the two pairs of critical quantities.

In the case of invariant electric charge $Q$, from critical pressure
Eq.(\ref{eq27}) we know that
\begin{equation}
\label{eq69} \frac{1}{l_c }= 4\left( {\frac{\pi A(x-2)}{n(n-1)x}}
\right)^{1/2}\left( {\frac{2A}{(x-1)xB}} \right)^{1/(x-2)}.
\end{equation}
Substituting Eq.(\ref{eq69}) into the critical temperature
Eq.(\ref{eq55}) derived from the case of invariant cosmological
constant $l$, one can get
\[
T_c =\tilde {A}D^{-1/2} +\tilde {B}D^{1/2} +\tilde {C}\frac{\tilde
{A}'+\tilde {B}'D}{\tilde {C}'}D^{-1/2} =\frac{k(n-2)b^{-2\gamma
}(x-2)}{2\pi (1-2\gamma )(x-1)}D^{-1/2}
\]
\begin{equation}
\label{eq70} =\frac{k(n-2)(x-2)}{2\pi (1-2\gamma )l(x-1)}\left(
{\frac{2n(n-1)}{k(n-2)(x-2)}} \right)^{1/2}
=\frac{2A(x-2)}{(x-1)}\left( {\frac{2A}{(x-1)xB}} \right)^{1/(x-2)}.
\end{equation}
From Eq.(\ref{eq70}) and (\ref{eq27}) we can find that the two pair
of critical temperatures are the same. Because the critical
temperature, critical pressure and critical volume in
Eq.(\ref{eq27}) are all dependent on $B$, and $B$ is the function of
electric charge of black hole, the critical quantities should be the
function of electric charge of black hole. The critical temperature
and critical electric charge in Eq.(\ref{eq54}) are the function of
$l$. The relations between the both quantities are given by
Eq.(\ref{eq69}), thus critical pressure and critical volume can also
be expressed as the function of cosmological constant. The critical
electric potential from Eq.(\ref{eq54})
\begin{equation}
\label{eq71} U_c =\frac{1}{\lambda }E^{1/2}b^{(3-n)\gamma
}D^{-\lambda /(2-4\gamma )}=\frac{1}{\lambda }\left(
{\frac{k(n-2)}{2(x-1)}} \right)^{1/2}.
\end{equation}
From Eq.(\ref{eq71}), the critical electric potential is dependent
on the spacetime dimension $n$ and dilaton field $\gamma $ and is
independent of the cosmological constant $l$.

From above we find that when the relation (\ref{eq69}) is satisfied
for the charged topological dilaton AdS black hole the phase
transition like van der Waals vapor-liquid one will turn up.

The critical temperature, critical pressure, critical volume and
critical electric potential are given by Eq.(\ref{eq27}) and
(\ref{eq54}). The critical exponents are the same as the ones in van
der Waals vapor-liquid phase transition.

Because of the relation (\ref{eq69}), according to the critical
temperature, critical pressure and critical volume derived in Sec.3
we can obtain the critical electric charge and critical electric
potential.

From Eq.(\ref{eq69})
\begin{equation}
\label{eq72} B=\frac{2A}{(x-1)x}\left( {\frac{16l_c^2 \pi
A(x-2)}{n(n-1)x}} \right)^{(x-2)/2}.
\end{equation}
Substituting Eq.(\ref{eq71}) into the above equation, one can get
\begin{equation}
\label{eq73} q_c^2 =\frac{k(n-2)}{2(x-1)}b^{2(n-3)\gamma }D^{\lambda
/(1-2\gamma )}.
\end{equation}
which agrees with Eq.(\ref{eq55}). From Eq.(\ref{eq23}) and
(\ref{eq27}), the horizon of the black hole correspondent to the
critical points is
\begin{equation}
\label{eq74} r_{+c} =\left( {\frac{(x-1)xB}{2A}} \right)^{\lambda
/[(x-2)(1-2\gamma )]}\left( {\frac{(n-1)(1-\gamma )b^{-2\gamma
}}{4}} \right)^{\lambda /(1-2\gamma )}.
\end{equation}
Substituting Eq.(\ref{eq73}) into (2.13), one can obtain the
consistent critical electric potential with Eq.(\ref{eq71}). Thus
for the $(n+1)$-dimensional charged topological dilaton AdS black
hole, in Sec.3 and Sec.4 we can both derive the critical
temperature, critical volume, critical pressure, critical electric
potential and critical electric charge. For the two pairs of
parameters $(P-V)$ or $(Q-U)$, the phase structure and critical
phenomena of black hole are the same as the ones in van der Waals
vapor-liquid system.

Due to $A\propto k$, the critical temperature, critical pressure,
critical electric potential and critical electric charge derived
above will tend to zero, however the critical volume will tend to
infinity. Therefore when $k=0$, for the $(n+1)$-dimensional charged
topological dilaton AdS black hole there no exist similar phase
transition to the one in the van der Waals system.

When $k=-1$, the results are complicated. The critical
temperature$T_c <0$, critical electric charge and electric potential
are imaginary numbers. $v_c \propto (-1)^{1/(x-2)}$, $P_c \propto
-(-1)^{2/(x-2)}$, which is dependent on the value of $x$. Whether
this process can happen or not? If happens, what physical mechanism
it should correspond? These questions should be studied further.

In this paper we studied the phase structure and critical phenomena
of the $(n+1)$-dimensional charged topological dilaton AdS black
holes. We find that the phase structure in the canonical ensemble
significantly depends on the parameter $k$, dimensionality $n$,
dilaton field $\gamma $ and cosmological constant $l$ or the
electric charge $q$ of black hole. We consider Hawking temperature,
electric charge, the cosmological constant, electric potential and
volume as the state parameters and analyzed the phase structure and
the critical phenomena. We do not discuss the effects of dilaton
field $\gamma $ and $b$ in Eq.(\ref{eq6}) on the phase structure and
the critical phenomena and will leave this for further work.

\bigskip

\section*{Acknowledgements}
This work is supported by NSFC under Grant
Nos.(11247261;11175109;11075098;11205097).

\end{document}